\documentclass[preprint,12pt]{elsarticle}
\usepackage{amssymb}

\newcommand{\ba}{\begin{eqnarray}}
\newcommand{\ea}{\end{eqnarray}}
\newcommand{\beqs}{\begin{eqnarray}}
\newcommand{\eeqs}{\end{eqnarray}}

\begin{document}

\begin{frontmatter}

\title{ 
Statistical and systematical errors in analyses of separate
			experimental data sets in high energy physics}

\author{  R. Orava$^a$,
O. V. Selyugin$^b$ \footnote{ e-mail: selugin@theor.jinr.ru  }
}
\address{
 \sl {$^a$Helsinki
Institute of Physics and University of Helsinki,} \centerline{\sl
PL64, FI-00014 University of Helsinki, Finland,  } \\
{\sl  now at CERN, CH-1211 Geneva 23, Switzerland }   \\
 {\sl $^b$BLTPh, JINR, Dubna, Russia}
}

\begin{abstract}

Different ways of extracting parameters of interest from combined data sets of separate experiments are
investigated accounting for the systematic errors. It is shown, that the frequentist approach may yield
larger $\chi^2$ values when compared to the Bayesian approach, where the systematic errors have a Gaussian distributed prior calculated in quadrature. The former leads to a better estimation of the parameters.
 A maximum-likelihood method, applied to different "gedanken" and real LHC data, is presented.
 The results allow to choose an optimal approach for obtaining the fit based model parameters.

\end{abstract}


\end{frontmatter}


\section{Introduction}

To select a decent model, a physicist has to account for a multitude of experimental data sets registered
 during different beam conditions and in varying detector set-ups, exhibiting vastly different statistical
  and systematical errors. For believable testing of theoretical models, the systematic uncertainties should
  be under control \cite{Hudson,Bityukov11,Knoet}. Frequency analyses, based on the likelihood ratio and other methods,
  are widely in use in particle physics \cite{Prosper,Erler15} and in high-energy astrophysics.
   When averages of different experimental results for the same quantities are computed,
   each one including both the statistical and systematical errors, the combined error of these is usually referred to.

Both sources of uncertainty constitute important pieces of information,
since the statistical errors usually scale in proportion to the sample size,
while this is not the case for the systematical or theoretical sources of uncertainty.
 The systematical error cannot be reduced by simply increasing the statistical significance
 of separate experimental data points \cite{Prosper,Erler15}.

Experimental measurements are still  sometimes  presented without inclusion of their systematical uncertainties,
and it is not always obvious whether the quoted over-all error bars include both statistical
 and systematical uncertainty.
  In fact, the actual background rates and shapes of the measured distributions are sensitive
 to a number of experimental quantities, such as calibration constants, detector geometries,
  poorly known material budgets within experiments, particle identification efficiencies etc.
   A 'systematical error', referred to by a high energy physicist, usually corresponds
    to a 'nuisance parameter' by a statistician.

The uncertainties, due to  propagation of imperfect knowledge of nuisance parameters
that cannot be constrained by the same data set, lead to systematical uncertainties.
 However, the uncertainties that are purely related to the fit, are referred to as statistical uncertainties.
 The uncertainties due to   calculations, such  as uncertainty propagation
  and  treatment of systematical effects, have to be accounted for, as well,
  since the conventional statistics does not guarantee consistent treatment of these,
   but rather an {\it ad hoc} procedure is typically used  \cite{DAug1,DAug3}.

There are two fundamentally different ways of including statistical
 and systematical errors in the fitting procedure.
 The first one, mostly used in connection with the differential cross sections,
 takes into account the square of the statistical and systematical errors in quadrature:
    $\sigma_{tot}^2 = \sigma_{stat.}^2+ \sigma_{syst.}^2$.
   The second approach accounts for the basic property of systematical errors,
   i.e. the fact that these errors have the same sign and size in proportion to the effect
    they have in another set of the same experimental data.
    To account for these properties, extra normalization coefficients for the measured data
     are introduced in the fit. For simplicity, this normalization is often transferred
     into the model parametrization, while it - in reality - accounts for the unknown normalization
      of the experimental data. This method is often used by research collaborations to extract,
      for example, the parton distribution functions of nucleons
        \cite{Stump01,exmp1-26,exmp1} and nuclei \cite{EPPS16})
       in high energy accelerator experiments, or in astroparticle physics  \cite{Koh15}.

There are number of studies addressing the way to include systematical errors
in experimental measurements (see, for example, references  \cite{Bityukov11,Fichet},   
 and references therein).
 In these studies, a predefined region of allowed values is usually considered,
 in order to define the magnitude of the signal above a large background.
 This differs from the cases where a number of different experimental data sets
 are spread over intervals that are specific to each experiment.
 The systematic errors, in this case, will have many different contributions.
  For example, the TOTEM Collaboration presented eight different sources of systematic errors
  in their analysis of reference    \cite{T7a}.   
  The signs of these systematic effects may vary,
   but usually there is a single dominating systematic uncertainty present.
   At the high energy accelerators, it is often   the machine luminosity error, that plays the main role.

The luminosity error has the same sign for the whole data set collected by an experiment.
When using the square sum approach to evaluate the over-all error in the fit,
the sign constraint is lost. Due to this problem, several additional normalization coefficients
 are introduced to account for the systematic errors in accelerator based physics
 \cite{Stump03,Sel-PRD15},   or in cosmology  \cite{Koh15,Ankowski16}.  
   Both methods can also be used simultaneously, by accounting
   for the bulk of the systematical errors by the square sum method and,
    in addition, for the maximum one, as a nuisance parameter in the fit.

Sometimes a more complicated combination of the two methods is used   \cite{Erler15,Ghosh17}. 
 In reference   \cite{Ge12}, 
 for example, the total  $\chi^2$ is separated into three parts:
       $\chi^2= \chi^2_{para}+\chi^2_{sys}+\chi^2_{stat}$
 and each term is estimated separately.
 There are systematic uncertainties of different origins to be addressed in theory computations  \cite{Charles16}. 
  Here, the experimental systematic uncertainties that have the same sign for a set of experimental data are considered.

In the second part of the present analysis, the two ways of accounting for the statistical
 and  systematical errors of different data sets are discussed. In the third part,
  the simplest linear model, similar to a toy model discussed in reference  \cite{Barlow-17}, 
   is analyzed. In the fourth and fifth parts, a more complicated nonlinear models,
   tested against four separate simulated data sets, is addressed.
   In the sixth section, an analysis of five sets of actual experimental LHC data is presented.
    Finally, in seventh part,  the analysis of some practice using the systematic errors is made.
    In Conclusions, the summed results are presented.

\section{ 
  Error combination}

The data sets provided by individual experiments are unique to each experimental set-up,
 and can be considered as statistically independent.
 These data sets are then used to fit a model, using a number of model parameters of interest,
  or nuisance parameters are introduced to account for possible uncertainties
  in normalization of data to account for the varying experimental conditions.

In the frequentist approach, the most widely used goodness-of-fit statistics
in hypothesis testing is  $\chi^2$,
the value of which is determined by the residual between the fitted model
and the data, using no input from the prior knowledge.
 Thus,      $\chi^{2}_{min} = \chi^{2} (\alpha_{j})$   
 represent the goodness-of-fit statistics for the minimum,
   $\chi^2$ solution for $\alpha_{j}$.  

The likelihood can be written in a "binwise" form,
i.e. in the form that accounts for the choice of bin widths.
The effect of choosing the bins can be modeled by shifting the signal
and background templates up and down, corresponding to the degree of uncertainty
\ba
 \mathcal L (\vec{n}|\vec{s}, \vec{b},\mu,\beta) = \left[ \prod_{i=1}^{n_{bins}} Pr(\hat{E}_{i}|F(a_{j},\delta)) \right]
 \ea
Here $\hat{E}_{i}$ is the observed event number and  $F(a_{j},\delta)$
the value resulting from a version of the model
 where the model parameters  $\alpha_{j}$ and nuisance parameter $\delta$
 were used. In case a sufficiently large number of model parameters
  are used, the Gaussian prior can be assumed for the distribution of experimental data,
   and the likelihood becomes
   \ba
 \mathcal L (\vec{n}|\vec{s}, \vec{b},\mu,\beta) =
  \prod_{i=1}^{n_{bins}} \frac{1}{\sqrt{2\pi} \sigma_{i}} e^{-(\hat{E}_{i}-F(a_{j},\delta))^2/2\sigma_{i}^{2}}
 \ea

  According to the frequentist approach, for a model with the correct dependence
 on its parameters of interest, moving the parameters to their "true"
  values means that the corresponding likelihood attains its maximum value.
  This procedure is equivalent to the minimization of the value of the likelihood $\chi^2$
   \ba
 -2 ln \mathcal L (x_{i}; \mu, \sigma) = \sum_{i}^{n} \frac{(\hat{E}_{i} - \mu)^2}{\sigma_{i}^{2} } + n (ln 2\pi + 2 ln \sigma ),
 \ea
  where the last term does not impact the position of the minimum of $\chi^2$.
  This term, however,  impacts the absolute size of $\chi^2$ at the location of the minimum.

For determining the parameters of interest, only location of the minimum $\chi^2$ is required.
Minimization of  $-2 ln \mathcal L (x_{i}; \mu, \sigma)$  can proceed either analytically
or numerically by finding the zeros of the first derivative with respect to $\mu$ and $\sigma^{2}$.
 The following maximum-likelihood estimates $\mu$ and $\sigma^{2}$ are obtained:
  $\hat{\mu}=[\sum_{i}^{n}(x_{i})]/n$ and
  $\sigma^{2} = [\sum_{i}^{n}(x_{i} - \hat{\mu})^{2}]/n$.

The maximum- likelihood estimate of $\sigma^{2}$  is biased, in the sense that its average value deviates from the true  $\sigma^{2}$.
In the following, for simplicity, all statistical errors are assumed to be of the same order of magnitude, $\sigma_{i}=\sigma_{st.}$.
 The systematic error is a nuisance parameter reflecting
 the detection efficiencies or uncertainty in measuring the luminosity.
  This error can be accounted for as a bias within the model
  where the corresponding errors $\sigma_{\delta}$ are adopted.
  Assuming the Gaussian prior for such a bias, the likelihood becomes
    \ba
 \mathcal L (\vec{n}|\vec{s}, \vec{b},\mu,\beta) =
   \int_{-\infty}^{\infty} \frac{1}{\sqrt{2\pi} \sigma_{st.}} e^{-(\hat{E}_{i}-(F(a_{j})-\delta))^2 /2\sigma_{st.}^{2}}
  \frac{1}{\sqrt{2\pi} \sigma_{syst.}}  e^{-\delta^2/2 \sigma_{syst.}^{2} } d \delta
 \ea
 The integration has a standard representation, for example in reference  \cite{DAug2}  
 it is of the form
   \ba
 \mathcal L (\vec{n}|\vec{s}, \vec{b},\mu,\beta) =
  \frac{1}{\sqrt{2\pi} \sqrt{\sigma_{st.}^{2}+\sigma_{syst.}^{2} }    } e^{-(\hat{E}_{i}-F(a_{j}))^2 /2(\sigma_{st.}^{2}+\sigma_{syst.}^{2}) }.
 \ea

 The total error is now expressed in terms of the sum of squares:
  $$\sigma_{tot}=\sqrt{\sigma_{st.}^{2}+\sigma_{syst.}^{2}}$$.
  It should be noted that this result assumes the Gaussian form for the bias.
In this case, the systematical errors will also have their signs distributed according
 to the Gaussian form. This contradicts the assumption that all signs
  of the systematic errors  of one origin have the same sign for a chosen set of experimental data.

In the following, to compare possible sizes of $\chi^2$
 in case of the squared errors, and for fitting additional normalization coefficients,
  the statistical and systematic errors are assumed to be of equal size.
  In the case of the squared errors,  $\chi^2$  can be simply written as
\begin{eqnarray}
\chi^{2}=\sum_{i=1}^{n} \frac{ ( \hat{E}_{i}  - F_{i}(a_{j}) )^2  }
{\sigma_{i-st.}^{2}+\sigma_{i-syst.}^2 }
    \label{eq6}
 \end{eqnarray}

 Assuming, that all the errors are of the same size, and that  $F_{i}=\bar{x}$. 
 then $\sigma_{tot}^{2} =2 \sigma^{2}$, $\sigma = 1/\sqrt{N}$  and  $\chi^2$  becomes
\begin{eqnarray}
\chi^{2}=\sum_{i=1}^{n} \frac{ ( \hat{E}_{i}  - \bar{x} )^2  }
{2 \sigma^{2} } = \frac{1}{2 \sigma^{2} } \sum_{i=1}^{n} \hat{E}_{i} - \frac{n \bar{x}^2 }{ 2 \sigma^{2} } \\ \nonumber
 = \frac{N}{2 } \sum_{i=1}^{n} \hat{E}_{i} - \frac{n N \bar{x}^2 }{ 2 }   = A_{1} - A_{2}.
    \label{eq8}
 \end{eqnarray}
  As a result, difference of the two terms appears in equation (7). 
When the systematic errors are taken into account as an additional normalization coefficient,
 $k$, and the size of this coefficient is assumed to have a standard error,
   $k=1\pm \sigma $,  
 \begin{eqnarray}
\chi^{2} &=&\sum_{i=1}^{n} \frac{ ( k \hat{E}_{i}  - \bar{x} )^2  }{2 \sigma^{2} }
 + \frac{(1-f_k)^2}{\delta^{2}_{i(norm)}} \\ \nonumber
  &=& \frac{1\pm\sigma}{2 \sigma^{2} } \sum_{i=1}^{n} \hat{E}_{i} - \frac{n \bar{x}^2 }{ 2 \sigma^{2} }
 \pm\frac{(1-f_k)^2}{\sigma^{2}}
 = B_{1} - B_{2} \pm \Delta.
    \label{exmp1}
 \end{eqnarray}
    The last term, $(1-f_k)^{2} / \delta^{2}_{i(norm)}$ is small compared to the others.
 Although this term could be of significance in model fits to the data, it is neglected in the following.

For large $N$ , the difference in $\chi^2$ is
	  \begin{eqnarray}
  \Delta \chi^{2}= (A_{1}-B_{1}) + (B_{2} - A_{2}),
    \label{exmp1}
 \end{eqnarray} 								
which can be written as
    \begin{eqnarray}
  \Delta_{1} \chi^{2}= \frac{1}{\sigma^2} (\frac{1}{2}-f_{k}^{2})) \sum_{i=1}^{n} \hat{E}_{i}
    \label{exmp1}
 \end{eqnarray}
and
   \begin{eqnarray}
  \Delta_{2} \chi^{2}= \frac{1}{\sigma^2} (1-\frac{1}{2})) n \bar{x}^2.
    \label{exmp1}
 \end{eqnarray}

 If the set of experimental data has no bias, then
 $f_{k}=1$ and $\sum_{i=1}^{n} \hat{E}_{i} - n \bar{x}^2   \approx 0$,
  In case the set of data is biased, then
  $\sum_{i=1}^{n} \hat{E}_{i} - n \bar{x}^2   > 0$  and $\Delta \chi^2 < 0$.
  Hence, the $\chi^2$  for the squared errors will be smaller than in the case
   where the additional normalization for the set of experimental data is accounted for.
   This will be revisited below in examples where simulated data sets are used.

\section{Model ab}
To study the influence of additional normalization coefficients in model fitting, a simple model
  ($\rho(t) =a + b t$), 
   also analyzed in Ref.[24], for simulated "experimental" data is considered, based on two sets of data.
   The first data set is constrained to be within the $t$-interval from
    $t =0.5$ to $12.5$ with $\Delta t=0.5$. 
   The second data set is constrained to
 $t=8$ to $20$ with $\Delta t=0.5$; 
   hence the two data sets have $50$ points.

As the initial values of the model parameters $a$ and $b$, $a = 0$ and $b = 1$ are chosen.
To simulate the $50$ points of "experimental" data, a random procedure with $10\%$ statistical errors is used
 (see the Appendix). To account for possible systematical errors, the second set is shifted by $20\%$
 with respect to the initial (simulated) values. As a result, two variants of the "experimental" data is obtained:
  the first one with zero systematic errors, and the second one with the $+20\%$ systematic error.
 A fit is then performed to the two data sets to determine the crucial model parameters
   on the basis of the experimental data.

   The first model variant for the experimental data set without
systematical errors is considered first. The results are summed up
by the first row in Table 1. The $\chi^2$ value obtained is small, and
the size of parameter $a$ remains practically zero. Parameter $b$ has a
value close to its "true" value.

Next, the simulated data with the
assumed $+20\%$ systematic errors is considered. The fitting procedure
is carried out for the following three cases: (1) accounting for
the statistical errors, only; (2) the errors are assumed to have
the form $\sigma_{tot}^2 = \sigma_{stat.}^2+ \sigma_{syst.}^2$;
 (3) $\sigma_{tot}^2 = \sigma_{stat.}^2$,
 where the systematic errors are included in fitting the extra normalization coefficients.
  The second, third and fourth rows of Table 1 list the results of the case assuming $+20\%$ systematical errors.
  The minimum $\chi^2$ is obtained for the cases assigned with the squared statistical and systematic errors.
   The magnitudes of the model parameters have, however, sizable deviations
    from their true values and large errors.
    In Figure 1, it can be seen that the best fit is obtained for the case
     where an additional normalization is included in the fit.

\begin{table}
 \caption{Description of Model ab, $\rho(t) = a+ b t$ (syst.er. = stat. er.) 
  with shift $n_i=1.$ } 
\label{Table-1}
\begin{center}
\begin{tabular}{|c|c|c|c|c|} \hline
 Model         & $\sum_{N} \chi^{2}$      & $a$     &  $b$      &  $n_i$      \\  \hline
   $\sigma_{st.}^2$      & $38.65$    & $-0.0056\pm0.04$ & $0.968\pm0.016$ &   $1.;1_{fix.}$      \\ \hline
     $\sigma_{st.}^2$      & $81.1$    & $-0.115\pm0.04$ & $1.08\pm0.02$ &   $1.;;1_{fix.}$      \\
   $\sigma_{st.}^2 + \sigma_{syst.}^2$    &$3.9.$   &  $-0.17\pm0.4$ & $1.22 \pm0.09$ &   $1.;1_{fix.}$.    \\
    $\sigma_{st.}^2$      & $27.4$    & $-0.006\pm0.03$ & $1.02\pm0.02$ &   $0.945;;1.19$      \\
%
%
 \hline
\end{tabular}
\end{center}
  \end{table}

\begin{figure}
    \includegraphics[width=.8\textwidth]{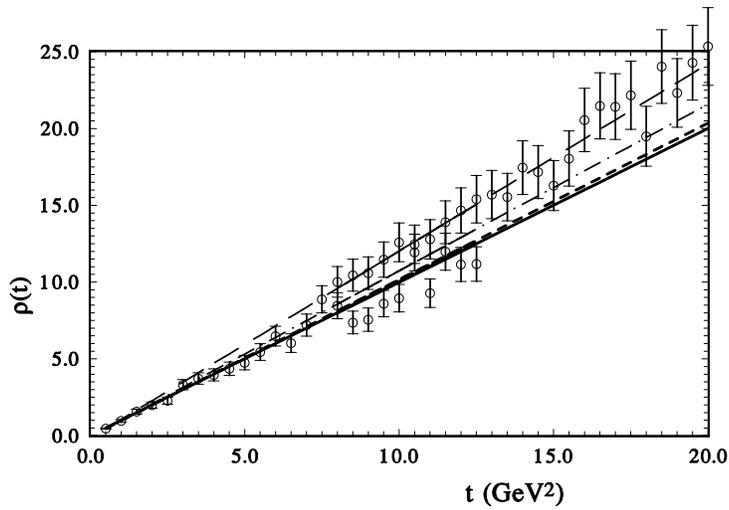}
\vspace{1cm}
  \caption{ Linear fit,  $\rho(t)= a+b t$, 
   to the simulated data with $+ 20\%$ systematic errors
(the second set of simulated "experimental" data).
(1) Dash-point lines indicate the calculation accounting for the statistical errors, only;
(2) long dashed line indicates the calculation with  $\sigma_{tot}^2 = \sigma_{stat.}^2+ \sigma_{syst.}^2$;
(3) short dashed line indicates the calculations where  $\sigma_{tot}^2 = \sigma_{stat.}^2$ 
and extra normalization coefficients are used;
 (4) solid line indicates the exact calculation  $\rho(t) = t$.  
   }
\end{figure}


\section{Model A-Gd-1}

Next, experimental data is emulated by using the familiar expression
\begin{eqnarray}
dS_{0}/dt= 1/(1. + \sqrt{t}/0.71).
    \label{exmp1}
 \end{eqnarray}

In Equation (12), the parameters determined by our "experimental" data are exactly known.
The following calculations are restricted to the $t$-region of $0 < t < 20$
for the $200$ simulated experimental data points with an assumed bin width of  $\Delta t=0.01$. 


The simulated experimental points are calculated for four $t$-intervals,
  $t=0-5$, $5-10$, $10-15$ and $15-20$.  
  The statistical and systematical errors are assumed to be
      $1\%, 2\%, 4\%, 8\%$, 
      for the four $t$-intervals, respectively.
  A random procedure (see the Appendix) is then applied that accounts for the statistical errors.
   As a result, an unbiased simulated  data set is obtained for  $dS_{1}/dt $. 

The standard fit to the simulated data was done by using the FUMILI program \cite{FUMILI,FUMILY}. 
This is preferred instead of the commonly used MINUIT \cite{MINUIT} 
which includes three separate minimization methods,
and may lead to results that have intrinsic dependence
on the different representations used in simulating the experimental data.

Next, the following model parametrization with free parameters is used to fit the simulated data:
	\begin{eqnarray}
dS/dt= h/(1. + t^{\alpha}/L)
    \label{M1d}
 \end{eqnarray}								
The results are listed in Table 2. It is clear, that despite of the large difference
 of the $\chi^2$ values, the fit parameters attain the same sizes in both cases,
  where either only statistical errors or the sum of the squared systematical
  and statistical errors are considered. The sizes of the fitting parameters
  are very close to the parameter values used in calculation of the simulated data.

A bias is then introduced for a separate data set, by assigning systematic errors for each data interval
$n_{i}=1.01, 0.98, 1.04, 0.92 $.
    \begin{eqnarray}
dS_{i}/dt= n_{i} h/(1. + t^{\alpha}/L).
    \label{M1d}
 \end{eqnarray}									
As a result, a modified simulated data set is defined having different bias for each data interval, $dS_{n}/dt $. 
 Obviously, the sign of the systematic error is the same for every point of each $t$-interval.

\begin{table}[b]
 \caption{Description of Model A-Gd: $dS_{1}/dt= h/(1. + t^{\alpha}/L)$ ($\sigma_{syst.} = \sigma_{stat.}$ with shift $n_i=1.$ }
\label{Table-2}
\begin{center}
\begin{tabular}{|c|c|c|c|c|} \hline
%
 Model         & $\sum_{N} \chi^{2}$      & $h$     &  $\alpha$   & $L$  
     \\  \hline
      &         &               &             &       \\
   $\sigma_{st.}^2$      & $297.7$    & $0.966\pm0.02$ & $0.521\pm0.006$ & $0.771\pm0.03$  f      \\
   $\sigma_{st.}^2 + \sigma_{syst.}^2$    &$148.9.$   &  $0.966\pm0.026$ & $0.52 \pm0.008$ & $0.771\pm0.04$
     \\
   &         &              &                &         \\
 \hline
\end{tabular}
\end{center}
  \end{table}

In the first two cases, in Tables 2 and 3, symmetric distribution of the signs of the systematic errors is assumed.
The sign is freely distributed according to the Poisson or Gaussian form;
 also a non-symmetric distribution of the signs of the systematic errors is considered.

The model fit was done for three different cases, where:  \\
	a) Only systematical errors were taken into account  $\sigma_{tot}^2 = \sigma_{st.}^2$  ; \\
	b) The systematical and statistical errors were squared:  $\sigma_{tot}^2 = \sigma_{st.}^2 + \sigma_{syst.}^2$ \\
	c)  $\sigma_{tot}^2 = \sigma_{st.}^2$  and $n_i$  were taken into account as nuisance parameters in the fit.

The results are presented in Table 3. The $\chi^2$ value is smaller in the case of the squared errors,
  $\sigma_{tot}^2=    \sigma_{st.}^2 +  \sigma_{syst.}^2 $.
   It is four time smaller when compared to case $c)$,  where only statistical errors,
   $\sigma_{tot}^2=    \sigma_{st.}^2  $ are accounted for, and extra normalization coefficients are used as free parameters.

\begin{table}
 \caption{Description of Model A-Gd-8:  $dS_{n}/dt= h/(1. + t/L)^{\alpha}$   ($\sigma_{syst.} = \sigma_{stat.}$)
    with shift  $n_i=1.01; 0.98; 1.04; 0.92$ }
\label{Table-3}
\begin{center}
\begin{tabular}{|c|c|c|c|c|c|} \hline
%
 Model         & $\sum_{N} \chi^{2}$      & $h$     &  $\alpha$   & $L$    &  $n_i$      \\  \hline
      &         &               &             &   &       \\
   $\sigma_{st.}^2$      & $320.$    & $0.968\pm0.02$ & $0.52\pm0.006$ & $0.767\pm0.03$    &  $n_{i}=1.$      \\
   $\sigma_{st.}^2 + \sigma_{syst.}^2$    &$37.6.$   &  $0.94\pm0.07$ & $0.53 \pm0.02$ & $0.808\pm0.1$ &  $n_{i}=1.$            \\
  $\sigma_{st.}^2$  & $154.9$   &  $0.97\pm0.04$ & $0.49 \pm0.01$ & $0.69\pm0.001$     &  $n_{i}$       \\
   &         &              &                &       &    \\
 \hline
\end{tabular}
\end{center}
  \end{table}

  The basic objective in this analysis is not to find the maximum
likelihood of the fit, but to determine the true sizes of the model
parameters.
 Obviously, the third case with its extra free nuisance parameters introduced for normalization, would give the best technical fit result. It should be noted that constant $h$ stays practically the same for the first and third cases.
 In fact, this results from the assumption of symmetric distributions of the signs of the systematical errors.

Consider next the asymmetric case.
For this, the bias for the separate sets of simulation data is assumed to be given as
  $n_{i}= 1.01,0.98, 1.04$ and $1.08$.   
The fit results for this case are shown in Table 4.
The $\chi^2$ of the squared errors is smaller than in the previous symmetric case.
 However, the sizes of the obtained parameters
 deviate more with respect to their true values.
 It is interesting to note, that for the last model variant, with the extra free normalization parameters,
  the resulting parameter values are clearly  closer to their true values when compared to the ones obtained in the symmetric case.

  The results do not change significantly, when the statistical and systematical errors
are increased, and allowed to change faster
 with increasing $t$ (for example, by increasing $t$ in steps of
$4\%, 8\%, 12\%, 16\%$ ).
The statistical and systematical errors are here assumed to have the same values.
The results are shown in Table 5 for the symmetric case, and in Table 6
for the asymmetric case.
Note that in these cases, the $\chi^2$ values
 for the squared errors and for the case with free normalization parameters,
 are very close to each other. However, the parameter values appear
  to be closer to their true values for the last model variant for both symmetric and asymmetric cases.

\begin{table}
 \caption{Description of Model A-Gd:  $dS/dt= h/(1. + t^{\alpha}/L)$   ($\sigma_{syst.} = \sigma_{stat.}$) in the non-symmetric case
(with bias $n_i=1.01; 0.98; 1.04; 1.08$) }
\label{Table-4}
\begin{center}
\begin{tabular}{|c|c|c|c|c|c|} \hline
%
 Model         & $\sum_{N} \chi^{2}$      & $h$     &  $\alpha$   & $L$    &  $n_i$      \\  \hline
      &         &               &             &   &       \\
   $\sigma_{st.}^2$      & $299.$    & $1.041\pm0.02$ & $0.495\pm0.006$ & $0.673\pm0.03$    & $n_{i}=1.$       \\
   $\sigma_{st.}^2 + \sigma_{syst.}^2$    &$30.2.$   &  $1.12\pm0.07$ & $0.47 \pm0.02$ & $0.59\pm0.1$ & $n_{i}=1.$         \\
  $\sigma_{st.}^2$  & $139.1$   &  $1.015\pm0.03$ & $0.493 \pm0.001$ & $0.686\pm0.002$     &$n_{i}$        \\
   &         &              &                &       &    \\
 \hline
\end{tabular}
\end{center}
  \end{table}


\begin{table}
 \caption{Description of Model A-Gd:  $dS/dt= h/(1. + t^{\alpha}/L)$ ($\sigma_{syst.} = \sigma_{stat.}$) 
  with the shift    $n_i=1.04; 0.92; 1.08; 0.84$ }
\label{Table-5}
\begin{center}
\begin{tabular}{|c|c|c|c|c|c|} \hline
%
 Model         & $\sum_{N} \chi^{2}$      & $h$     &  $\alpha$   & $L$    &  $n_i$      \\  \hline
      &         &               &             &   &       \\
   $\sigma_{st.}^2$      & $356.4$    & $0.83\pm0.04$ & $0.63\pm0.02$ & $1.12\pm0.11$    & $n_{i}=1.$       \\
   $\sigma_{st.}^2 + \sigma_{syst.}^2$    &$178.2.$   &  $0.83\pm0.06$ & $0.63 \pm0.03$ & $1.12\pm0.16$ &$n_{i}=1.$     \\
  $\sigma_{st.}^2$  & $177.2$   &  $0.99\pm0.2$ & $0.47 \pm0.03$ & $0.61\pm0.09$     & $n_{i}$        \\
   &         &              &                &       &    \\
 \hline
\end{tabular}
\end{center}
  \end{table}

\begin{table}
 \caption{ Description of Model A-Gd-Up:  $dS/dt= h/(1. + t^{\alpha}/L)$ ($\sigma_{syst.} = \sigma_{stat.}$)
  with the shift $n_i=1.04; 0.92; 1.12; 1.16$ }
\label{Table-6}
\begin{center}
\begin{tabular}{|c|c|c|c|c|c|} \hline
%
 Model         & $\sum_{N} \chi^{2}$      & $h$     &  $\alpha$   & $L$    &  $n_i$      \\  \hline
      &         &               &             &   &       \\
   $\sigma_{st.}^2$      & $322.8$    & $1.23\pm0.12$ & $0.476\pm0.02$ & $0.535\pm0.08$    & $n_{i}=1.$       \\
   $\sigma_{st.}^2 + \sigma_{syst.}^2$    &$161.4.$   &  $1.23\pm0.17$ & $0.475 \pm0.03$ & $0.535\pm0.12$ & $n_{i}=1.$     \\
  $\sigma_{st.}^2$  & $158.2$   &  $1.06\pm0.13$ & $0.46 \pm0.04$ & $0.57\pm0.14$     &$n_{i}$        \\
   &         &              &                &       &    \\
 \hline
\end{tabular}
\end{center}
  \end{table}



\section{Model A-Gd8-1}

A more complicated model is examined below. For this, experimental
data is simulated by using the following formula that is close to
the observed differential cross sections, and is proportional to
the fourth power of the proton dipole form factor
\begin{eqnarray}
dS/dt= h/(1. + t/L)^{8}
    \label{M8d}
 \end{eqnarray}
 Here the parameters are chosen as $h = 100$ and $L = 0.71$.
As in the previous cases, the entire $t$-interval from $t = 0 - 20$ is considered and
$200$ "experimental" points separated into four intervals are generated.
 The statistical errors are assumed to be  $2\%; 4\%, 8\%$, and $12\%$;
 the systematic errors as $4\%; 8\%, 16\%$, and $24\%$.
 Random procedure is then used to generate four sets of simulated experimental data.
 Supposing that the true form of the data is not known,
  an exponential model is adopted to describe of the generated data.
   In terms of combined exponentials
\begin{eqnarray}
dS/dt= h_{1} exp[-\alpha_{1} t) +h_{2} exp[-\alpha_{2} t).
    \label{M2e}
 \end{eqnarray}
To determine the optimum parameters for the model (16),
the simulated data is assigned with small $1\%$ statistical errors and with zero systematical errors.
 A fit using four free parameters with these relatively small errors, yields an optimum $\chi^2$,
  given the dipole model (15), and very large $\chi^2$, value in case of the exponential model (16) (see Table 7).
   The parameters obtained for the dipole model well coincide with the parameters used
    in simulating the "experimental" data sample. For the exponential model,
    the parameters determined by the fit can be considered as the best description
    of this particular choice of the model.

In the following, the simulated data is assigned with both statistical and
 systematical errors and, as in the previous simple cases, the symmetric
 and asymmetric cases are investigated separately.
 The bias in the last $t$-interval is assumed to be $\mp 24\%$.

 The results for the symmetric case are shown in Table 8.
 Again, $\chi^2$ is sufficiently small for the model variant with squared
errors. The parameter values found are far off the ones obtained
when fitting with the assumed $1\%$ errors (Table 6, second row).
Despite of the larger $\chi^2$ values, the parameters (Table 8, third
row) obtained when including the extra normalization coefficients
better coincide with the true parameter values (Table 7, second
row). In the asymmetric case, the approach based on squared errors
yields better results (see Table 9), and the difference between the
two approaches (the one based on squared errors and the one based
on extra normalization coefficients) is less important when
compared to the symmetric case. Hence, in case the model is
sufficiently far from the reality, it is not obvious which model
variant to choose. The model variant with the extra normalization
parameters, however, wins over the square error case.

 In the following, the simulated data is analyzed by assuming statistical
errors of $2\%; 4\%, 8\%, 16\%$ and systematic errors $4\%; 8\%, 16\%, 24\%$
in terms of a true model.
 The fit is based on the model with three parameters
\begin{eqnarray}
dS/dt= h/(1. + t/L)^{\alpha}
    \label{M8fit}
 \end{eqnarray}

\begin{table}
 \caption{
 Description of Model A-Gd-8 $dS/dt= h/(1. + t/L)^{8}$
   by $h_{1} exp(-\alpha_{1} t) + h_{2} exp(-\alpha_{2} t)$
   ($\sigma_{st.} = 1 \% $)  $n_i=1.; 1.; 1.; 1.$ }
\label{Table-7}
\begin{center}
\begin{tabular}{|c|c|c|c|c|c|} \hline
%
 Model         & $\sum_{N} \chi^{2}$      & $h_{1}$     &  $\alpha_{1}$   & $h_{2}$    &  $\alpha_{2}$      \\  \hline
      &         &               &             &   &       \\
 Dipole    & $0.33$    & $99.99\pm1.1$ & $8.001\pm0.01$ & $0.71\pm0.002$    &  $ $      \\
 2 exp.     &$7027.$   &  ${\bf 74.04}\pm0.22$ & ${\bf8.64} \pm0.01$ & ${\bf2.68}\pm0.02$ &           ${\bf3.61}\pm0.004$    \\
   &         &              &                &       &    \\
 \hline
\end{tabular}
\end{center}
  \end{table}

\begin{table}
 \caption{Description of Model A-Gd-8 $dS/dt= h/(1. + t/L)^{8}$
   by $h_{1} exp(-\alpha_{1} t) + h_{2} exp(-\alpha_{2} t)$
    ($\sigma_{syst.} = \sigma_{stat.}$) symmetric case with bias $n_i=1.04; 0.92; 1.16; 0.76$ }
\label{Table-8}
\begin{center}
\begin{tabular}{|c|c|c|c|c|c|} \hline
%
 Model         & $\sum_{N} \chi^{2}$      & $h_{1}$     &  $\alpha_{1}$   & $h_{2}$    &  $\alpha_{2}$      \\  \hline
      &         &               &             &   &       \\
   $\sigma_{st.}^2$      & $1277 $    & $87.3\pm0.6$ & $9.82\pm0.05$ & $6.24\pm0.18$    &  $4.3\pm0.02 $      \\
   $\sigma_{st.}^2 + \sigma_{syst.}^2$    &$227  $   &  $87.3\pm1.3$ & $9.65 \pm0.1 $ & $5.2\pm0.3 $ &           $4.1\pm0.06$    \\
  $\sigma_{st.}^2$  & $862  $   &  $84.5 \pm3. $ & $10.3 \pm0.1$ & $8.7\pm0.5$     &  $4.4\pm0.1$       \\
   &         &              &                &       &    \\
 \hline
\end{tabular}
\end{center}
  \end{table}

\begin{table}
 \caption{Description of Model A-Gd-8 $dS/dt= h/(1. + t/L)^{8}$
   by $h_{1} exp(-\alpha_{1} t) + h_{2} exp(-\alpha_{2} t)$
   ($\sigma_{syst.} = \sigma_{stat.}$) asymmetric case with bias $n_i=1.04; 0.92; 1.16; 1.24$ }
\label{Table-9}
\begin{center}
\begin{tabular}{|c|c|c|c|c|c|} \hline
%
 Model         & $\sum_{N} \chi^{2}$      & $h_{1}$     &  $\alpha_{1}$   & $h_{2}$    &  $\alpha_{2}$      \\  \hline
      &         &               &             &   &       \\
   $\sigma_{st.}^2$      & $990$    & $86.5\pm0.6$ & $9.14\pm0.05$ & $2.97\pm0.09$    &  $3.6\pm0.03$      \\
   $\sigma_{st.}^2 + \sigma_{syst.}^2$    &$198$   &  $86.5\pm1.3$ & $9.24 \pm0.1$ & $2.1\pm0.2$ & $3.4\pm0.1$    \\
  $\sigma_{st.}^2$  & $836$   &  $88.4\pm2.6$ & $10.2 \pm0.1$ & $8.3\pm0.5$     &  $4.4\pm0.1$       \\
   &         &              &                &       &    \\
 \hline
\end{tabular}
\end{center}
  \end{table}

\begin{table}
 \caption{Description of Model A-Gd-8 by $dS/dt= h/(1. + t/L)^{\alpha}$
  ($\sigma_{syst.} = \sigma_{stat.}$) symmetric case with bias $n_i=1.04; 0.92; 1.16; 0.76$ }
\label{Table-10}
\begin{center}
\begin{tabular}{|c|c|c|c|c|} \hline
%
 $\sum_{N} \chi^{2}$      & $h$     &  $\alpha$   & $L$    &  $n_i$      \\  \hline
      &               &             &   &       \\
   $1211$  ($\sigma_{st.}^2$)   & $103.8\pm0.8$ & $8.6\pm0.06$ & $0.77\pm0.1$    & $n_{i}=1.$      \\
 $242$  ($\sigma_{st.+syst.}^2$)  &  $103.8\pm16$ & $8.6 \pm0.14$ & $0.77\pm0.2$ &$n_{i}=1.$   \\
    $616$   ($\sigma_{st.}^2$) &  $101.9\pm3.6$ & $7.8 \pm0.12$ & $0.69\pm0.02$     & $n_{i}$       \\
        &              &                &       &    \\
 \hline
\end{tabular}
\end{center}
  \end{table}

\begin{table}
 \caption{Description of Model A-Gd-8 by $dS/dt= h/(1. + t/L)^{\alpha}$
  ($\sigma_{syst.} = \sigma_{stat.}$) asymmetric case with bias $n_i=1.04; 0.92; 1.16; 1.24$ }
\label{Table-11}
\begin{center}
\begin{tabular}{|c|c|c|c|c|} \hline
%
  $\sum_{N} \chi^{2}$ (err.)     & $h$     &  $\alpha$   & $L$    &  $n_i$      \\  \hline
      &         &                       &   &       \\
     $1030$  ($\sigma_{st.}^2$)  & $109.4\pm0.8$ & $7.6\pm0.05$ & $0.65\pm0.01$    &  $n_{i}=1.$     \\
   $206$ ($\sigma_{st.+syst.}^2$)  &  $109.4\pm 2$ & $7.6 \pm0.11$ & $0.65\pm0.02$ &  $n_{i}=1.$    \\
  $576$ ($\sigma_{st.}^2$) &  $102.6\pm3.6$ & $7.8 \pm0.1$ & $0.69\pm0.02$     &$n_{i}$       \\
   &         &                    &       &    \\
 \hline
\end{tabular}
\end{center}
  \end{table}

The results for the symmetric and antisymmetric cases are shown in Tables 10 and 11.
 It is seen, that for the model variant based on using squared errors,
  $\chi^2$ is smaller in the symmetric case. Despite of the minimal $\chi^2$,
  the parameter values are far off the initial parameters defined for the model.
   Contrary to this, the model variant using extra free normalization coefficients,
   yields fit parameter values close to the true ones.

\section{Elastic  cross sections at the LHC}

 For testing the model hypotheses further, the data collected by the TOTEM and ATLAS Collaborations
 at $7$ and $8$ TeV is used below.
 At small four-momentum transfers, $-t$, in proton-proton elastic scattering processes
 close to the  diffraction peak region, there are five sets of experimental data used
 for measurements of the differential cross sections: two of them at $7$ TeV center-of-mass energy,
  $\sqrt{s}$, and three at $8$ TeV. The data sets come from different $t$-regions.
   The usual $ln (s)^2$ dependence of the total $pp$ cross section on c.m.s energy, is here assumed.
   As the very high LHC energies, the pre-asymptotic terms in the standard representation
   for the total cross sections can be safely neglected. For the description of the hadronic part
   of the differential cross sections, the standard exponential form of the hadron elastic scattering amplitude is taken
 \begin{eqnarray}
F_{h}(s,t)= i h  ln (s)^2 (1.- i\rho ) e^{B_{1}/2 t + B_{2}/2 t^2} G^2(t);
\label{Fh}
 \end{eqnarray}
 with the form factor	
\begin{eqnarray}
G(t) = \frac{4 m_p^2 - \mu t }{4 m_p^2-t}\frac{\Lambda^2}{(\Lambda - t)^2}.
\label{emff}
 \end{eqnarray}	
where $m_p$  is the proton rest mass,   $\Lambda=0.71$ GeV$^2$ and $\mu=2.79$.
 In calculations of the differential cross sections, the five spiral electromagnetic amplitudes
 and the Coulomb-hadron phase factor are accounted for (see, for example,  \cite{Sel-rho,HEGS0}). 
  At $7$ TeV,  the TOTEM measurements   \cite{T7a} 
 in the $t$-region of  $0.00515 <|t|<0.371$ GeV$^2$;
 and the ATLAS measurements \cite{ATL7} 
 in the region  $0.0062 <|t|<0.35$ GeV$^2$   are used.
  At $8$ TeV, the data published by the TOTEM Collaboration \cite{T8a} 
  in the $t$-regions of   $0.0285 <|t|<0.19$ GeV$^2$,   and   $0.000741 <|t|<0.191$ GeV$^2$
  and the data by the ATLAS Collaboration \cite{ATL8} 
  in the region of   $0.0105 <|t|<0.363$ GeV$^2$ are used.
   On the whole, these data sets contain $225$ data points.

 Some discrepancies exist in the total cross sections measured by the two Collaborations.
   From the separate analysis of each data set,
   the TOTEM Collaboration finds for the $pp$ total cross section:
   $\sigma_{tot} = 98.0 - 99.1 \pm 3  $ mbar at $7$ TeV and   $\sigma_{tot} = 101.7 \pm 2.9 $ mbar at $8$ TeV
    \cite{T7a,T8a}.
    The ATLAS Collaboration obtained somewhat smaller values:
    $\sigma_{tot} = 95.35 \pm 1.34  $ mbar at $7$ TeV
    and $\sigma_{tot} = 96.07 \pm 1.34 $ mbar at $8$ TeV \cite{ATL7,ATL8}.

All the above five data sets are analyzed below simultaneously.
The results of the analysis are listed in Table 12. In the first row (Table 12),
the result with only the statistical errors, and the additional normalization coefficients
with fixed by unity are shown. In the second row, the results of the same fitting procedure
 are shown but with the statistical and systematic errors
 ($\sigma^2 = \sigma_{st.}^2+\sigma_{syst.}^2$).
 Comparing these two results, it can be seen that the parameters of interest are practically
  the same, despite of the enormous difference in the over-all $\chi^2$.
  The total cross sections coincide with the ATLAS measurements.
  If the statistical errors are considered alone, but including the extra  normalization coefficients,
   the total $\chi^2$ decreases with respect to the first case  (Table 12, third row).
    The normalized TOTEM data lies above the ATLAS results.
    Note that this result is also obtained within the framework model of
      high energy general structure (HEGS) \cite{HEGS1,Diff16}. 

Here, a simple model parametrization of the hadronic amplitude is
used. Different forms of the amplitude should be considered and
their dependence on energy and four momentum transfer, while
accounting for the fit procedure.
 It is observed, that  the using approach based on the squared systematical
  and statistical errors, no  new results are obtained.
  Contrary to this, by including the systematic errors in model fitting,
   usage of extra normalization coefficient allows new results to be reached.
    The same conclusions were obtained above,  in connection of the model testing using the
     simulated "experimental" data samples.

\begin{table}
 \caption{Description of $d\sigma/dt$ at LHC energies
  }
\label{Table-1}
\begin{center}
\begin{tabular}{|l|c|c|c|c|c|c|} \hline
%
 $\sum_{N} \chi^{2}$; (err.)     & $h$     &  $B_{1}$ &  $B_{2}$ &$\rho$ &$\sigma_{tot}$ & $n_i$  \\
     &                       &                &       & &7{\small TeV}/$8${\small TeV} &  T;A;|T;T;A \\
 \hline
      &                  &             &   &    & &   \\
          $48337$  ($\sigma_{st.}^2$)  & $0.30$ & $0.55$ & $-0.39$ &$0._{b}$ &$95.3/98.2$  &  $1.;1.;|1.;1.;1.$      \\
   $421$  ($\sigma_{st.+syst.}^2$)  &  $0.30$ & $ 0.55$ & $-0.45$ &$0._{b}$& $95.1/98.0$ &           $1.;1.;|1.;1.;1.$    \\  \hline
    &                      &                &       & & &   \\
   $1812$  ($7$ {\small TeV})  &  $0.31$ & $0.58$ & $-0.26$ &$0._{b}$ &$96.7$ &  $1.03;0.98.;|$      \\
  ($\sigma_{st.}^2$)  ($8${\small TeV}) &      &  &  & & $99.7$ &  $1.06;1.06;0.94$       \\
   &         &                              &       & & &   \\
 \hline
\end{tabular}
\end{center}
  \end{table}

 \section{Notes concerning additional normalization of data}

Besides the standard use of systematic errors, either as the
squares of statistical and systematic errors summed together, or by
using additional normalization coefficients, other approaches have
been recently introduced in error analysis \cite{un-pdf,un-sig}. 
 In Ref.  \cite{un-sig} 
the following expression was used for  $\chi^2=\chi^2_{stat} +\chi^2_{scale}$  
\begin{eqnarray}
\chi^{2}_{stat} &=&\sum_{k=1}^{L}  \sum_{i_{k}} \frac{ ( \omega_{k}  \sigma_{inv,i_{k}} - \sigma_{inv} (C,{\cal{T}})_{i_{k}})^2 }
    {\omega^{2}_{k} \sigma^{2}_{i_{k}} }
    \label{primer1}
 \end{eqnarray}

 The authors note: "$ \sigma_{inv,i_{k}}$ is the $i_{k}$
 data point for invariant cross section having total uncertainty
 $\sigma_{i_{k}}$,, which is taken as the quadratic sum of statistical and systematical uncertainties
 of each data point if both are stated separately." All the experimental errors are,
  therefore, considered in the analysis. However, the authors state in addition:
  "For each data set we allow a re-scaling by a constant factor $\omega_{k}$".
  The size of the scale factor was chosen as: "the average size of the systematic uncertainties".
  Unfortunately, such a procedure leads to double counting of systematic errors.
  The authors use a normalization factor in the denominator when calculating the total error.
  However, the normalization factor, $\omega_{k}$, centers around unity: when it is less than unity,
  the total error decreases and vice versa, when it is above unity, the total error will increase,
   and the $\chi^2$ value tends to decrease. The additional term in $\chi^2$, expressed as
 \begin{eqnarray}
\chi^{2}_{scale} &=&\sum_{k=1}^{L}  \frac{ ( \omega_{k}-1)^2 }{ \sigma^{2}_{scale,k} }
    \label{primer2}
 \end{eqnarray}
will be independent of the sign of the term $ (\omega_{k}-1)$,
 and basically asymmetric properties of the $\chi^2$ are recovered.
   The authors in Reference \cite{un-sig} 
   end up to be mistaken in their approach to parameter fitting.

 \section{Conclusion}

All experimental data are associated with finite systematical errors.
 To reliably determine their sizes is of essential importance,
 and great care should be exercised in evaluating them.
 Erroneous treatment of the systematic errors can lead to fundamentally faulty conclusions
  when extracting model parameters through a fit.
  Different approaches in addressing the systematic errors,
  can lead to either right or wrong determination of the "true" model parameters,
   thereby influencing choice of a valid "true" model.
Complications in error calculation include propagation of uncertainties and treatment of systematic effects;
conventional statistical analyses do not usually involve consistent methods,
 but only {\it ad hoc} prescriptions to follow \cite{DAug1}. 
Present analysis shows that in model fitting, particularly in cases where
the systematical uncorrelated errors exceed the statistical errors,
additional normalization coefficients need to be introduced.


 In fact, when additional normalization coefficients are introduced in the fitting procedure,
  the $\chi^2$ values reached can end up being larger compared to the usage of the sum of squared errors.
  However, the parameters of interest of the tested model will be closer
   to their "true" values allowing to better validate the correct model description of experimental data.

\vspace*{0.5cm}
{\bf Acknowledgments}
 {\it The authors would like to thank J.-R. Cudell for fruitful discussions concerning the paper.}  \\

 \section{Appendix: Random procedure}

After model calculation of exact values using the simulated
"experimental" data, statistical and systematical distortions have
to be introduced. A procedure was developed to accomplish this, and
to account for possible small oscillations while avoiding external
bias. The available interval in $t$ is first divided into segments,
where each $t$-region has its own statistical and systematical
errors. For example, the first segment has $1\% *n$ statistical and
$2\% *(\pm n)$ systematical errors, where n is the order of this
interval of $t$. In this case, the random procedure for the
statistical errors is made and all the measurement points are moved
up by $2$ percent. The sign of the statistical error is calculated as
follows:

   The sign of the statistical error  is calculated as \\

$      znaks=0. $ \\
 $          k=i $   \\
 $       rad=dsqrt( (1.d0*k+0.021*k*k)/(1.+(.001d0*k+0.00021*k*k)))/sqrt(2.*k)  $ $		 z10=dmod(rad*10000000.,2.) $  \\ $          z1=int(z10)            $  \\
 $ 	 afn=modulo(z1,2.)             $  \\
 $          if(afn.eq.0.) then $       \\
 $          znak1=1.   $                \\
 $          else       $                \\
 $          znak1=-1.  $                \\
 $           end if    $                 \\

	$ 	z10=dmod(rad*1000000.,2.)   $        \\  $        z1=int(z10)            $     \\
 	$ afn=modulo(z1,2.)             $       \\
c          mod                            \\
c           afn=0.                         \\
   $       if(afn.eq.0.) then     $        \\
   $       znak2=-1.              $         \\
   $       else                   $         \\
   $       znak2=1.               $         \\
   $        end if                $          \\
   $ znak=znak1*znak2             $          \\
   $           znaks=znaks+znak   $           \\
%
Random numbers are first chosen by using the standard FORTRAN procedure, and a circle from 1 to 50. - \\
   {\bf call $random_number(u);  xfree=u(10) $} \\
%
%
 For $i$ point take some number \\
  {\bf   ax=(i+12)*.212 } \\
  and  {\bf   bx=(i+5)*.8 } \\
%
Then, using the system clock get the third number: \\
      {\bf   call $system_clock(count)
	          seed=count $}  \\
The numbers are then inputted in the standard Fortran procedure: \\
      {\bf   call $random_seed(put=seed); \\
	   call random_number(u)$  }

 $        x=(u(2)+u(10))/2.   $         \\
 $         x0=u(2)            $          \\
 $   y0=u(8)                  $          \\
%
  	$ y=(x+y0)/2.  $               \\
%
        Increase the errors by $n$ times $ en=8$.                  $        \\
   $   deri=eri(i)*en                                                          $         \\
   $                                                                           $         \\
   $    p0=ani(i)+znak* deri*dsqrt( -2.*dlog((x0+y)/2.))*dsqrt(1.-y0**2)       $         \\
  $   . *  (sin(2.*3.1415926*(y0+x)/2.)+cos(2.*3.1415926*(x0+y)/2.))/2.       $          \\
  $    p1=ani(i) +znak* deri*dsqrt( -2.*dlog((x0+y)/2.  ) )                   $           \\
  $   .    *2.*(x)/(0.5+y)                                                    $           \\
  $    p2=ani(i)+ znak*deri*dsqrt( -2.*dlog((y0+x)/2.  ) )                    $            \\
  $   .    *2.*(y)/(0.5+x)                                                    $           \\  
  $                                                                           $            \\
  $    p0a=ani(i)+znak* deri*sqrt( -2.*dlog((x+x0)/2.))*dsqrt(1.-y**2)        $            \\
  $   .    *2.*(y0)/(0.5+x0)                                                  $             \\
  $    p1a=ani(i) +znak* deri*dsqrt( -2.*dlog((y0+y)/2.    ) )                $             \\
  $     p2a=ani(i)+znak* deri*dsqrt( -2.*dlog( (y+y0)/2.  ) )                 $              \\
  $                                                                            $             \\
  $       rad1=(p0 +p1+p2+p0a+p1a)/5.                                          $              \\
  $       rad2=(p0 +p1+p2+p0a+p2a)/5.                                          $              \\
  $       rad3=(p0 +p1+p2+p1a+p2a)/5.                                          $              \\
  $       rad4=(p0 +p1+p0a+p1a+p2a)/5.                                         $               \\
  $       rad5=(p0 +p2+p0a+p1a+p2a)/5.                                         $              \\
  $       rad6=(p1 +p2+p0a+p1a+p2a)/5.                                         $               \\
   $     rad=(rad1+rad2+rad3+rad4+rad5+rad6)/6.                                                \\

$              ak=u(5)*100.   $          \\
$ak1= amod(ak,10.)            $           \\
$       ak2=(ak-ak1)/10.+1.   $           \\
$ak3=amod(ak2,2.)             $           \\
                                           \\
  $    ch0= ((p0-ani(i))/deri)**2   $        \\
  $    ch1= ((p1-ani(i))/deri)**2   $         \\
  $    ch2= ((p1-ani(i))/deri)**2   $         \\
 $     chs0=chs0+ch0       $                   \\
 $     chs1=chs1+ch1       $                    \\
 $     chs2=chs2+ch2       $                    \\
 $      an0=(p0-ani(i))/deri      $              \\
 $      an1=(p1-ani(i))/deri      $             \\
 $      an2=(p2-ani(i))/deri      $              \\
$	san0=san0+an0    $                            \\
$	san1=san1+an1    $                             \\
$	san2=san2+an2    $                             \\
%
%
%
%
%
%
%
%
%




\end{document}